\documentclass[namedreferences]{solarphysics}
\usepackage[hyperref,optionalrh]{spr-sola-addons} 

\usepackage{graphicx,color,breakurl} 

\newcommand{\etal}{{\it et al.}}
\newcommand{\eg}  {{\it e.g.}}

\newcommand{\degr}{\mbox{$^\circ$}}%
\newcommand{\dtm} {\mbox{$\delta\tau_\mathrm{mn}$}}%
\newcommand{\dtmn}{\mbox{$\delta\tau_\mathrm{mn}/N$}}%
\newcommand{\bm}  {\mbox{$|B|_\mathrm{mn}$}}%
\newcommand{\bave}{\mbox{$\langle |B| \rangle$}}%
\newcommand{\pdf}{\mbox{$\mathrm{pdf}(\lambda)$}}%

\begin{document}
\begin{article}
\begin{opening}

\title{Long-Term Variation of Helioseismic Far-Side Images and What Causes It}

\author[addressref={aff1},corref,email={junwei@sun.stanford.edu}]{\inits{J.}\fnm{Junwei}~\lnm{Zhao}}
\author[addressref={aff2}]{\fnm{Grace~Y.}~\lnm{Jiang}}
\author[addressref={aff1}]{\fnm{Ruizhu}~\lnm{Chen}}

\address[id=aff1]{W.~W.~Hansen Experimental Physics Laboratory, Stanford
           University, Stanford, CA 94305-4085, USA}
\address[id=aff2]{Lynbrook High School, San Jose, CA 95129, USA }

\runningauthor{J.~Zhao et al.}
\runningtitle{Long-Term Variations of Far-side Images}

\begin{abstract}
A new time--distance far-side imaging technique was recently developed by utilizing multiple multi-skip acoustic waves. 
The measurement procedure is applied to 11 years of Doppler observations from the {\it Solar Dynamics Observatory} / Helioseismic and Magnetic Imager, and over 8000 far-side images of the Sun have been obtained with a 12-hour temporal cadence. 
The mean travel-time shifts in these images unsurprisingly vary with the solar cycle.
However, the temporal variation does not show good correlations with the magnetic activity in their respective northern or southern hemisphere, but show very good anti-correlation with the global-scale magnetic activity. 
We investigate four possible causes of this travel-time variation. 
Our analysis demonstrates that the acoustic waves that are used for mapping the Sun's far side experience surface reflections around the globe, where they may interact with surface or near-surface magnetic field, and carry travel-time deficits with them. 
The mean far-side travel-time shifts from these acoustic waves therefore vary in phase with the Sun's magnetic activity.
\end{abstract}


\end{opening}

\section{Introduction}
\label{sec1}

Seeing active regions (ARs) on the Sun's far side is useful for monitoring developments of solar magnetic activity and forecasting arrivals of large ARs on the Sun's earth-facing side, and local-helioseismic techniques have made possible the detection of far-side ARs by using $p$-mode helioseismic waves. 
Helioseismic holography first succeeded in detecting a large AR near the center of the far side \citep{lin00}, and the method was later further developed to include the entire far side \citep{bra01} and to enhance the signal-to-noise ratio \citep{lin17}.  
A time--distance helioseismic technique, which utilized waves with three-, four-, and five-skip waves, was also developed and validated using numerical simulation \citep{zha07, har08, ilo09}. 
More recently, using {\it Solar Dynamics Observatory} / Helioseismic and Magnetic Imager (SDO/HMI: Scherrer et al. 2012, Schou et al. 2012) full-disk Doppler observations, \cite{zha19} completely re-developed their time--distance far-side imaging code and combined 14 measurement schemes that utilized three-, four-, five-, six-, and eight-skip waves, substantially enhancing the detection quality of the far-side ARs. 
The Sun's helioseismic far-side images were not just used to monitor the Sun's magnetic activity, but were also used to study the Sun's long-term variations. 
Analyzing the helioseismic-holography far-side images obtained using {\it Solar and Heliospheric Observatory}/Michelson Doppler Imager (SOHO/MDI: Scherrer et al., 1995) data, \cite{gon09} found that the mean helioseismic phase-shifts, deficits of which indicate far-side ARs, varied substantially with the phase of the solar cycle for Cycle 23. 
They attributed this long-term phase-shift variation to the Sun's seismic radius change with solar cycles.

In addition to the Sun's photometric radius measured in visible light, the Sun's radius can also be measured through $f$-mode oscillations on its surface \citep{sch97}. The radius measured this way is known as seismic radius $\mathrm{R_\odot}$. 
Later studies found that $\mathrm{R_\odot}$ changed with the phase of solar cycles, slightly larger during solar minima and smaller during maxima with a variation of a few kilometers \citep{lef05, kho08}. 
More recently, \cite{kos18} analyzed 21 years of combined observations by SOHO/MDI and SDO/HMI, and found that $\mathrm{R_\odot}$ reduces by $1-2$\,km during the activity maxima and the most significant radius change occurs at the depth of $5\pm2$\,Mm. 
Despite a completely different approach, the long-term far-side images that are primarily obtained from the low-degree deep-penetrating $p$-mode waves also indicate a variation of $\mathrm{R_\odot}$ \citep{gon09}, which seems consistent with those obtained by the global helioseismology analyses.

Having multiple sets of far-side images from different numbers of wave skips, the newly-developed time--distance far-side imaging method \citep{zha19} has an advantage over the helioseismic-holography images \citep{gon09} in evaluating whether the measured acoustic phase-shift (or travel-time shift for the time--distance helioseismic technique) variation is due to the Sun's $\mathrm{R_\odot}$ change or owing to other factors. 
In this article, we examine the long-term variation of the travel-time shifts in the time--distance far-side images measured from the {\it SDO}/HMI observations during Solar Cycle 24, and investigate the physical causes of the variation. We present our data-analysis procedure and results in Section 2, discuss the results in Section 3, and give conclusions in Section 4.

\section{Data Analysis and Results}
\label{sec2}

\subsection{Data}
\label{sec21}

\begin{figure}[!t]
\centerline{\includegraphics[width=0.85\textwidth]{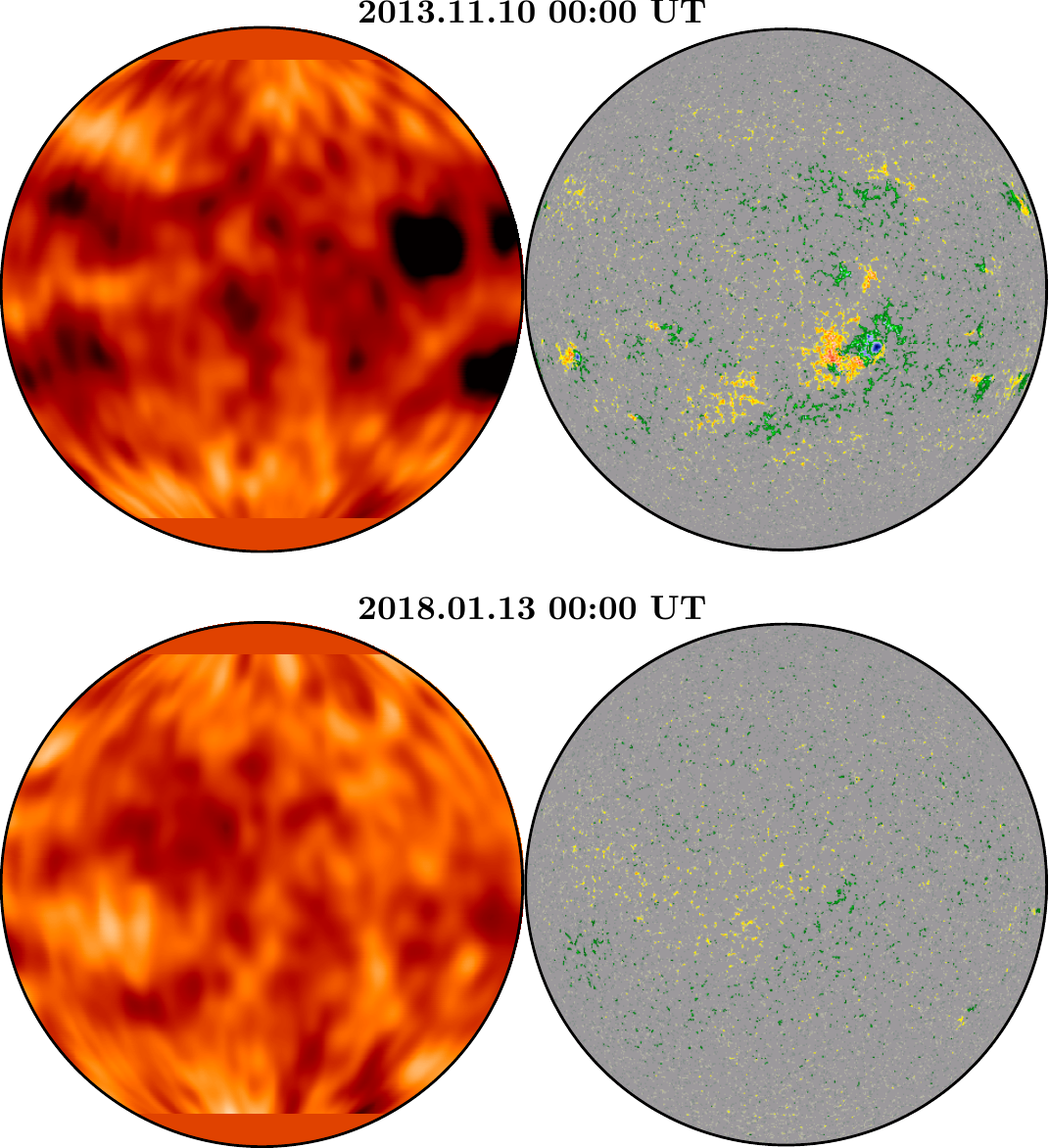} }
\caption{Examples of the Sun's helioseismic far-side images ({\it Left}) along with near-side magnetograms ({\it Right}) when both sides of the Sun were magnetically active ({\it Upper}) and quiet ({\it Lower}). 
Dark patches in the far-side images indicate ARs, and the blue-green (orange-red) patches in the magnetograms indicate positive (negative) magnetic regions.
The near-side magnetograms are plotted as the solar disk observed from near the Earth, and the far-side images are plotted as if the solar disk were observed from a far-side position on the extended Earth--Sun line.} 
\label{FS_examples}
\end{figure}

The time--distance helioseismic far-side imaging technique that was recently developed by \cite{zha19} is applied to the SDO/HMI data from 1 May 2010, the first observing day after the instrument's commissioning, through 30 April 2021 --  a total of 11 years. Each far-side image is obtained from a 31-hour observing period, and two far-side images are obtained every day with the middle time of the data period at 00:00 UT and 12:00 UT of the day. Figure~\ref{FS_examples} shows two examples of far-side images when the Sun was very active with large ARs on both the far and near sides, and when it was less active with no visible ARs on either side. The helioseismic far-side images are actually maps of the acoustic travel-time shifts measured relative to the Sun of activity minimum years (1 January 2018 -- 31 December 2019 for this study), and dark patches in such images (negative travel-time shifts) indicate where ARs are located. It is generally believed that the travel-time deficits corresponding to far-side ARs are due to the Wilson depression associated with magnetic regions \citep{lin10}.

For this time--distance far-side imaging method, each image is composed of and overlapped by 14 different sets of partial-far-side images made from waves experiencing different number of skips: three, four, five, six, and eight. 
In the time--distance diagram, there are two branches corresponding to each $n$-skip signal (refer to Figure~2 of Zhao \etal\ 2019), the lower one corresponding to helioseismic waves traveling less than $360\degr$ (hereafter, 1$\degr$ represents 1$\degr$ in great-circle distance on the solar surface) and the upper one corresponding to waves traveling more than $360\degr$. 
For a four-skip wave, \eg\ if the wave starts from the near side and is reflected twice before reaching the far-side target point, then is reflected twice before traveling back to the near side, this measurement scheme is called 2$\times$2-lower if the total travel distance of the wave is less than $360\degr$, or 2$\times$2-upper if greater than $360\degr$. 
Thus, the 14 sets of measurement schemes include: 2$\times$2-lower, 2$\times$2-upper, 3$\times$3-lower, 3$\times$3-upper, 4$\times$4-lower, and 4$\times$4-upper, primarily used to map the far-side central regions within $\approx48\degr$ of the far-side disk center; and 1$\times$2-lower, 1$\times$3-lower, 1$\times$3-upper, 2$\times$3-lower, 2$\times$3-upper, 2$\times$4-lower, 2$\times$6-lower, and 2$\times$6-upper, primarily used for mapping regions within $\approx45\degr$ of the far-side limbs. 
Because different measurement schemes have different noise levels, not all the data from these measurement schemes are used in this study.

\subsection{Mean Travel-Time Shifts and Mean Magnetic Field}
\label{sec22}

\begin{figure}[!t]
\centerline{\includegraphics[width=0.80\textwidth]{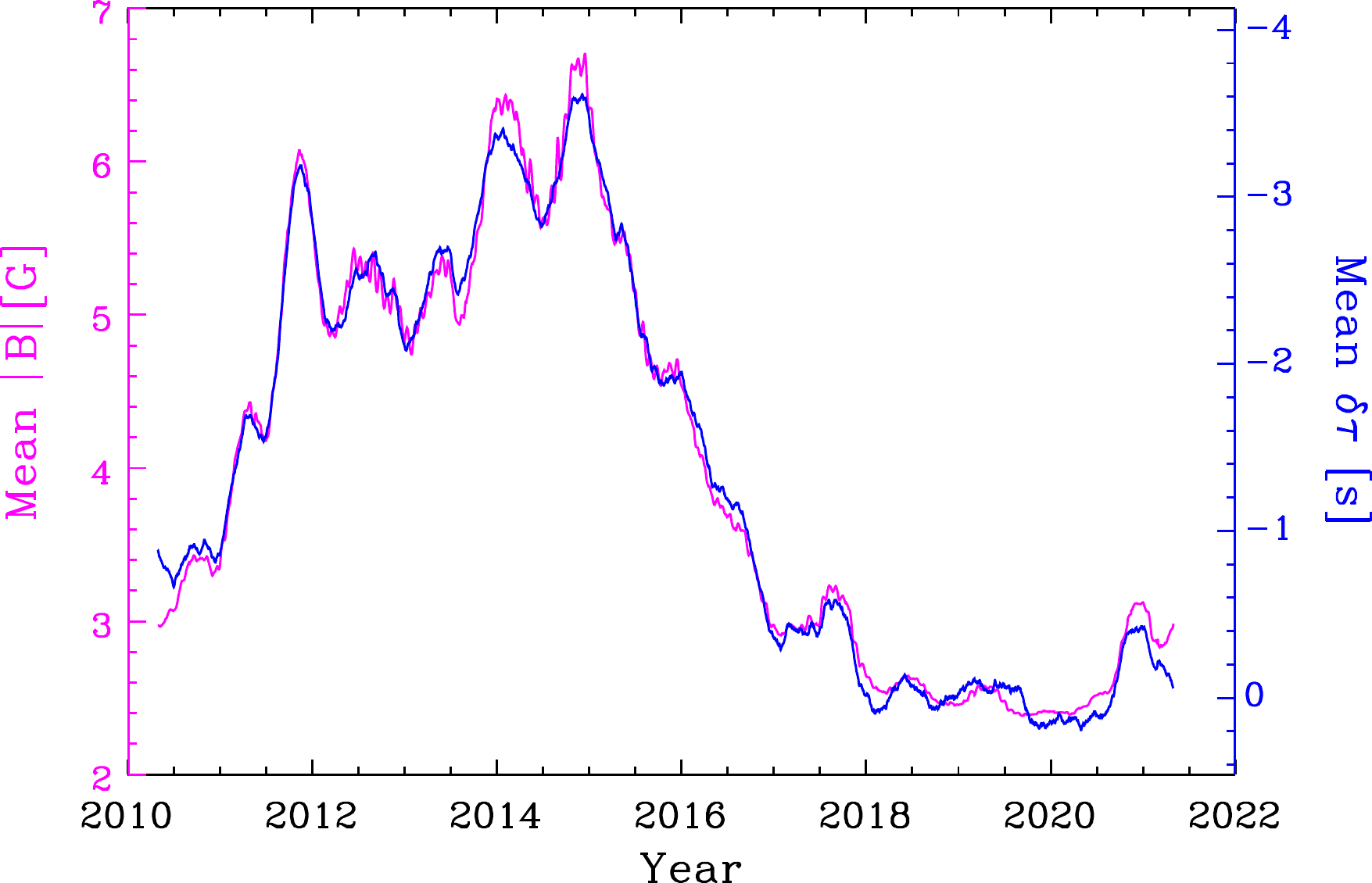} }
\caption{Long-term evolution of the Sun's near-side mean magnetic flux density [\bm] and far-side mean travel-time shifts [\dtm]. 
Note that for the axis for \dtm, values decrease upward.}
\label{long_time}
\end{figure}

To study the temporal evolution of the travel-time shifts in the time--distance far-side images during the 11-year period, we average the travel-time shift measurements throughout the entire far side that covers $180\degr$ in longitude and $120\degr$ in latitude in longitude--sin(latitude) coordinates, and we call the averaged values \dtm\ hereafter. 
Mean magnetic-flux density [\bm] is obtained daily by averaging the absolute values of the near-side full-disk magnetic field. Figure~\ref{long_time} shows the 120-day running averages of both \dtm\ and \bm, which show a very good anti-correlation, -0.995, between them.

\begin{figure}[!t]
\centerline{\includegraphics[width=0.70\textwidth]{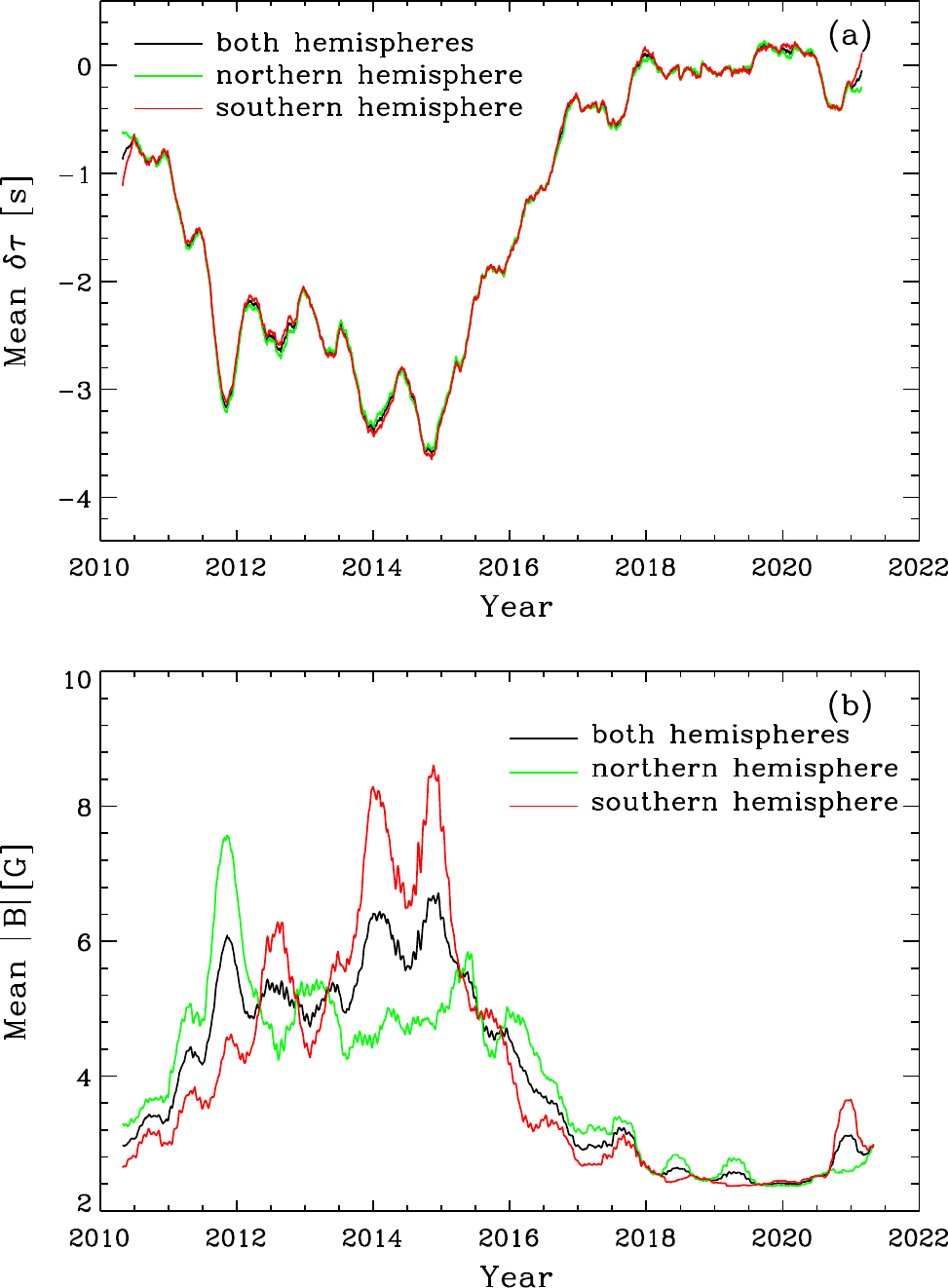}}
\caption{{\bf (a)} Long-term evolution of \dtm\ averaged from the Sun's far-side full disk as well as from both the far-side northern and southern hemispheres. 
{\bf (b)} Long-term evolution of the Sun's \bm\ obtained from the near-side full disk as well as from the near-side northern and southern hemispheres.}
\label{NS_hemi}
\end{figure}

It may not be surprising that \dtm\ has a very good anti-correlation with magnetic activity, because the travel-time deficits on far-side maps are mostly caused by the far-side magnetic regions, and the mean far-side magnetic-flux density is not expected to differ much from that of the near side, particularly after a 120-day average. 
To examine whether the contributions of far-side magnetic regions are the primary cause of the general variation trend of the \dtm, we plot in Figure~\ref{NS_hemi} the \dtm\ averaged from the Sun's entire far side, as well as those averaged from the far-side northern and southern hemispheres separately, together with the \bm\ from the entire near side as well as from the near-side northern and southern hemispheres. 
As can be seen, the \dtm\ from the far-side northern and southern hemispheres, as well as the entire far side, are nearly identical, while the magnetic-field strengths from the two near-side hemispheres show dramatic differences. 
This indicates that the general trend in the \dtm\ variation that is seen in Figure~\ref{long_time} is likely caused by a global-scale variation rather than by the magnetic-flux variation at local scales.

\subsection{Mean Travel-Time Shifts and Number of Skips}
\label{sec23}

The results shown in Figure~\ref{long_time} are essentially consistent with those reported by \cite{gon09}; however, the time--distance far-side imaging technique has one advantage over theirs in that we have multiple multi-skip measurements that provide us more data to track down the causes of the general variation trend of the \dtm. 
Comparing the \dtm\ measured from different measurement schemes, i.e. different combinations of wave-skips, may shed a light on how we can understand the \dtm\-variation with the solar cycle. 
Because the \dtm\ measured in the far-side central area has different properties from that measured in the far-side limb area, we analyze these measurements separately.

\begin{figure}[!t]
\centerline{\includegraphics[width=0.70\textwidth]{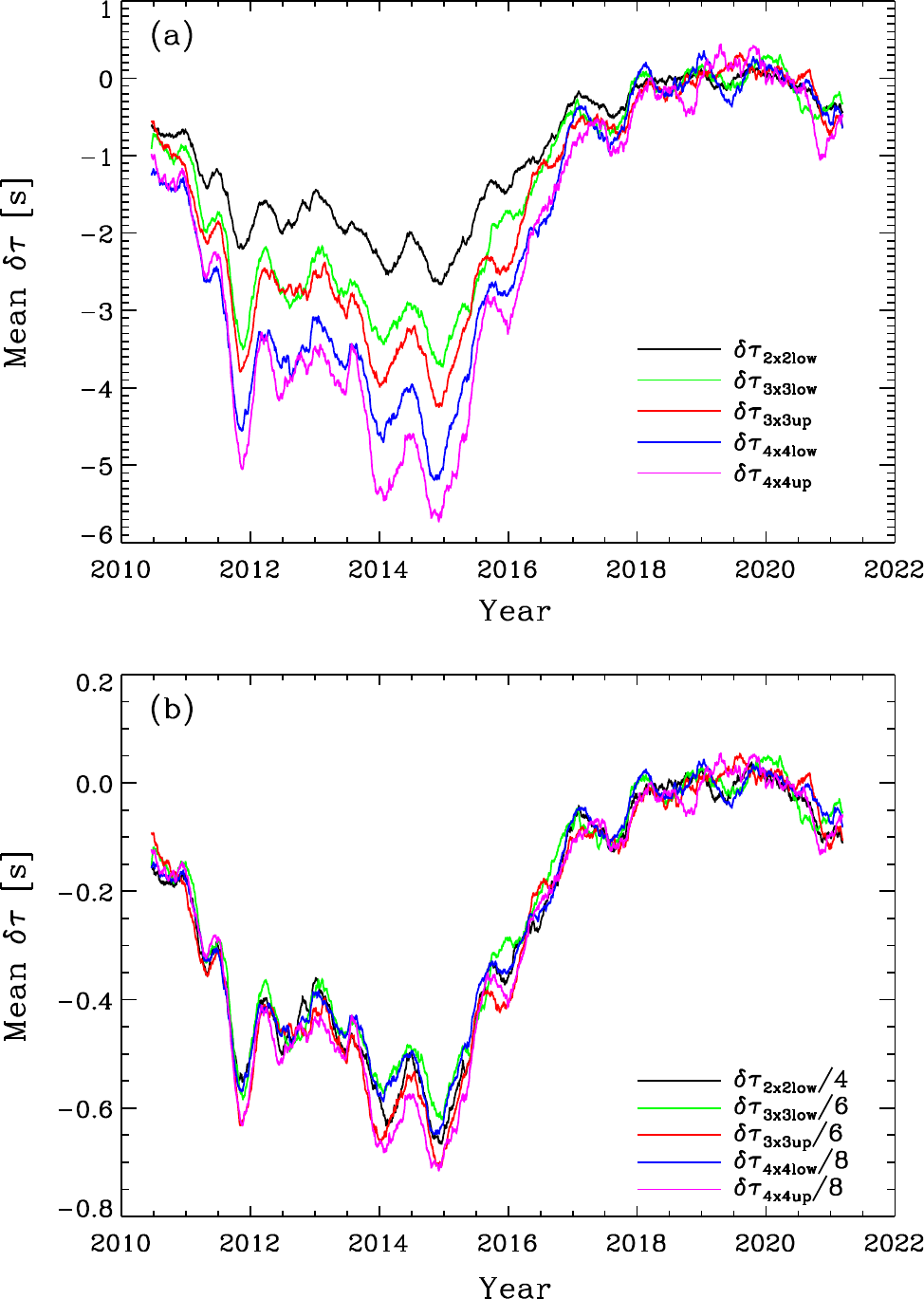} }
\caption{{\bf (a)} Temporal evolution of \dtm\ obtained from the far-side central areas using the measurement schemes marked in the panel. 
{\bf (b)} Same \dtm\ as in Panel {\bf a} but displayed after dividing their respective number of wave skips, i.e. \dtmn.
Note the measurements from the 2$\times$2-upper scheme are noisier and are not shown, but their basic trends are consistent with the ones shown.} 
\label{time_central}
\end{figure}

Figure~\ref{time_central}a shows the temporal evolution of \dtm\ measured from the far-side central area using the measurement schemes of 2$\times$2-lower, 3$\times$3-lower, 3$\times$3-upper, 4$\times$4-lower, and 4$\times$4-upper; Figure~\ref{time_central}b shows \dtmn -- \dtm\ divided by their respective number of skips used in the measurements. 
As can be seen, the \dtm\ from different measurement schemes show substantial differences in values although all of them show a very similar general trend of variations with the solar cycle; however, the \dtmn\ are essentially consistent with each other, indicating that a single skip from each measurement scheme gets a similar amount of travel-time reduction. 
Meanwhile, we want to emphasize that despite the close similarities among the \dtmn\ for different measurement schemes, the small differences between them are not negligible, and these important differences will be discussed later in Section~\ref{sec33}.

\begin{figure}[!t]
\centerline{\includegraphics[width=0.70\textwidth]{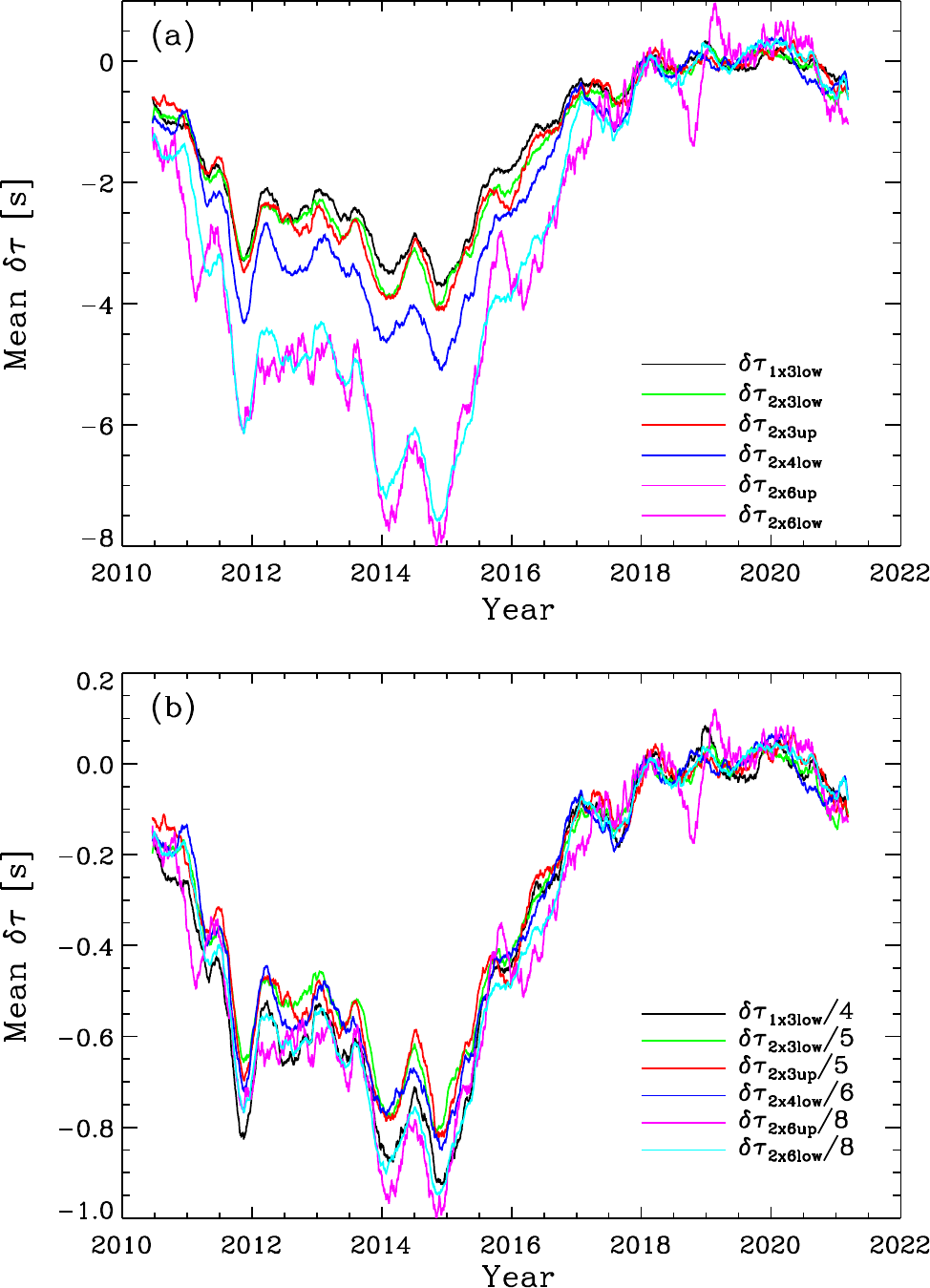} }
\caption{{\bf (a)} Temporal evolution of \dtm\ obtained from the far-side limb areas using the measurement schemes marked in the panel. 
{\bf (b)} Same \dtm\ as in Panel {\bf a} but displayed after dividing their respective number of wave skips, i.e., \dtmn.
Note that the measurements from the 1$\times$2-lower and 1$\times$3-upper are noisier and are not shown here, but their basic trends are consistent with the ones shown.} 
\label{time_limb}
\end{figure}

Similarly, Figure~\ref{time_limb} shows the \dtm\ and \dtmn\ measured from the limb areas of the far side, including the measurement schemes of 1$\times$3-lower, 2$\times$3-lower, 2$\times$3-upper, 2$\times$4-lower, 2$\times$6-upper, and 2$\times$6-lower. 
Similar to the result from the far-side central region, the \dtm\ from different measurement schemes differ substantially but the \dtmn\ are quite consistent with each other, again with non-negligible differences. 

\subsection{Mean Travel-Time Shifts and Latitude}
\label{sec24}

\begin{figure}[!t]
\centerline{\includegraphics[width=0.70\textwidth]{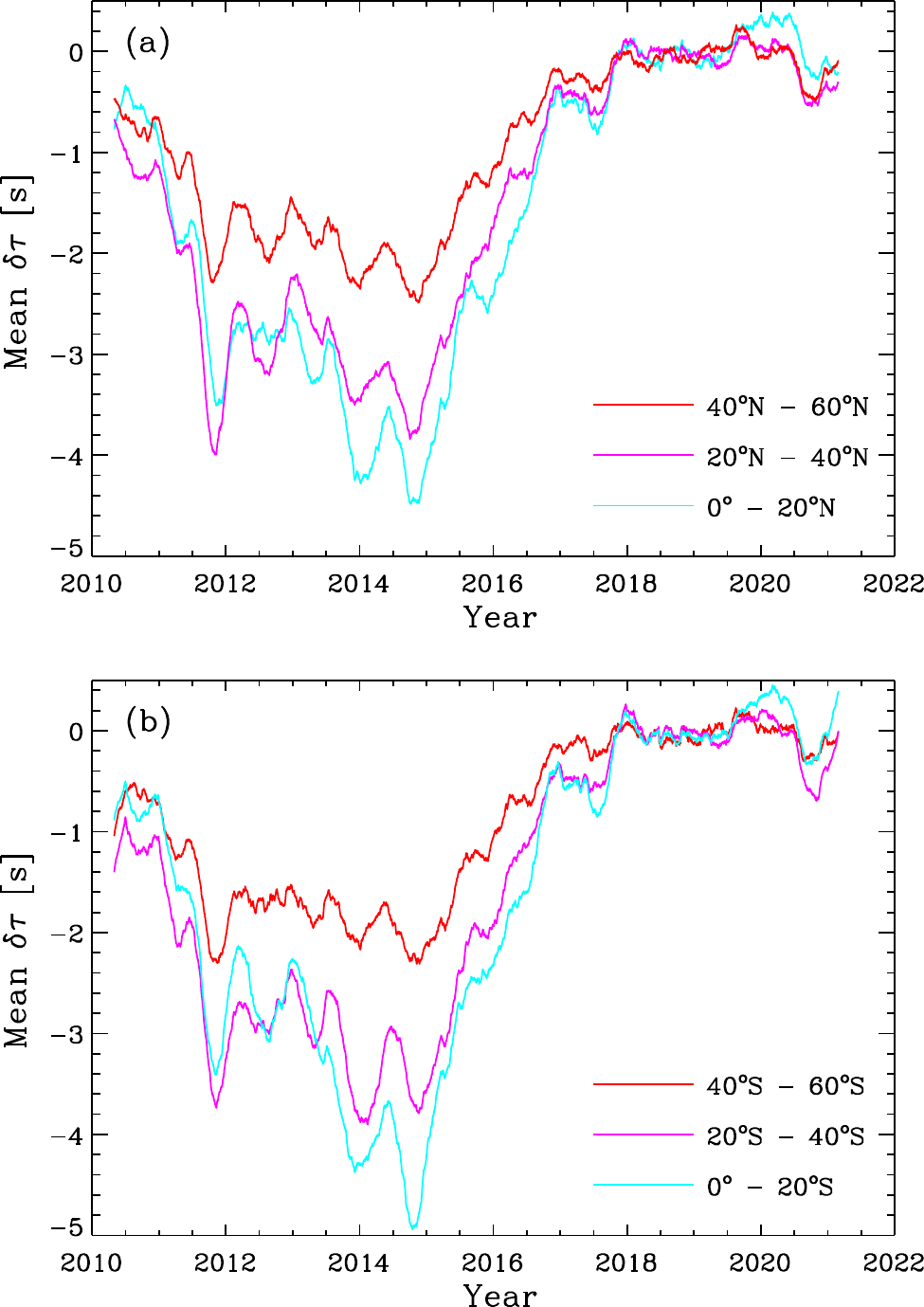} }
\caption{Temporal variations of far-side \dtm\ for selected latitudinal bands in the northern {\bf (a)} and southern {\bf (b)} hemispheres.
These data are from the final far-side images after combining all the measurement schemes.} 
\label{lat_var}
\end{figure}

Figure~\ref{NS_hemi} shows variations of \dtm\ for both the northern and southern hemispheres, and Figures~\ref{time_central}b and \ref{time_limb}b show variations of the \dtmn\ for different measurement schemes. 
All of these results exhibit very similar trends and amplitudes of variations. However, as shown in Figure~\ref{lat_var}, the variation trends of the \dtm, obtained from a few selected latitudinal bands in the combined far-side images (rather than images of different measurement schemes), show significantly different amounts of travel-time shifts, despite that they still show similar variation trends. 
The \dtm\ in the lower-latitude bands have greater magnitudes than those in the higher-latitude bands. 
These results indicate that although the general trend of the \dtm-variations seems to be caused by a global factor, the variation amplitudes are actually disk-location dependent.

\section{Discussion}
\label{sec3}

\subsection{What Causes the Long-Term \dtm-Variation?}
\label{sec31}

The \dtm-variations shown in Section~\ref{sec2} are essentially consistent with those by \cite{gon09}, except that our results with different numbers of wave skips provide more information that can help us better interpret the results. 
It may not be surprising to see that the far-side \dtm\ varies with the phase of solar cycles, however, determining what is the physics behind these measured variations may not be that straightforward. 

\begin{figure}[!t]
\centerline{\includegraphics[width=0.85\textwidth]{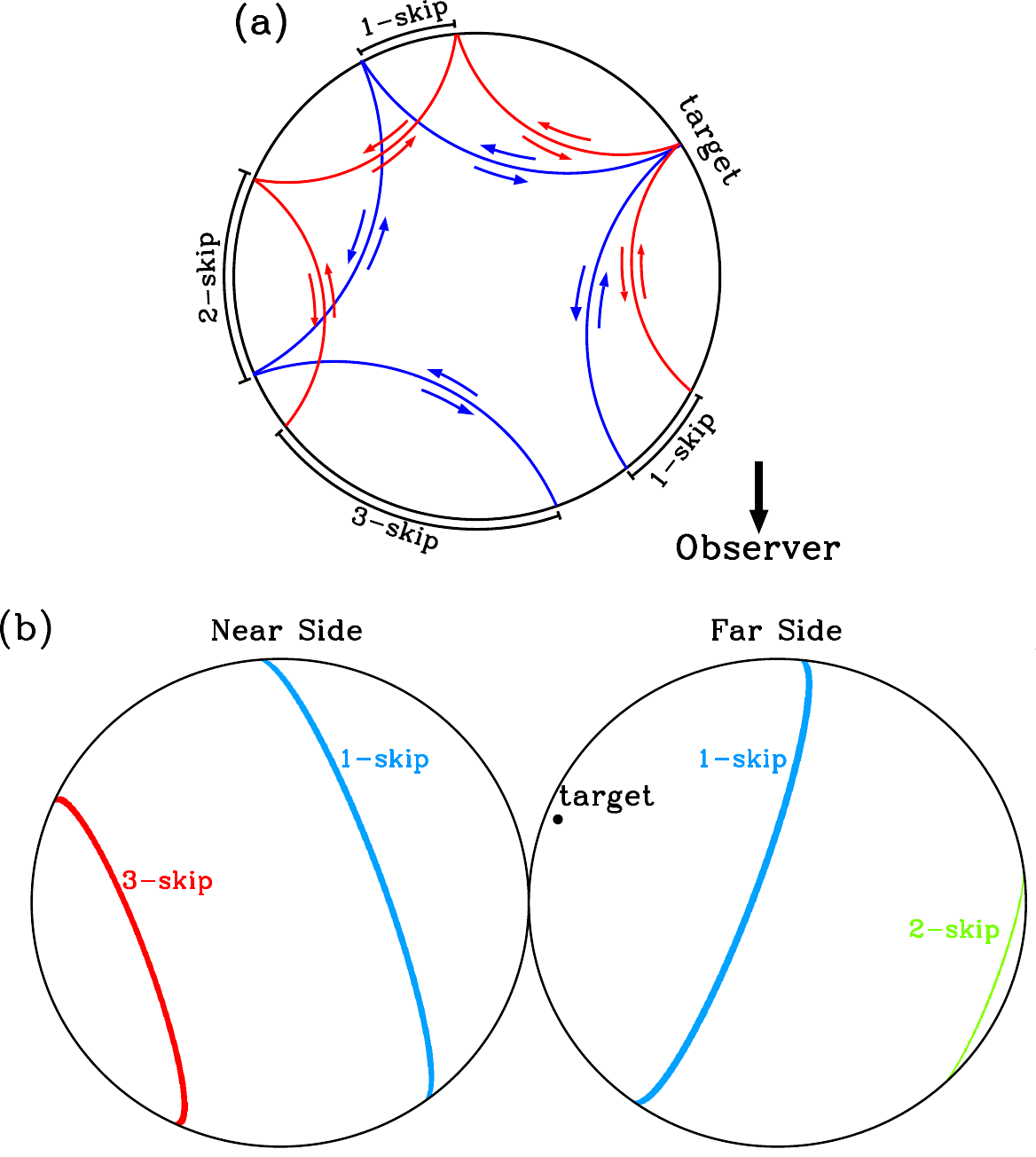} }
\caption{{\bf (a)} Great-circle slice showing the measurement scheme for 1$\times$3-lower case. 
Observations inside the one-skip and three-skip partial-annuli on the near side, i.e. lower half of the circle, are used for calculating travel-time shifts at the far-side target. 
{\bf (b)} An example showing the surface-reflection locations, in both near- and far-side disks, corresponding to a far-side target at a longitude of $15\degr$ past the west limb and latitude of $20\degr$N. 
Single-skip distance for this example is $75\degr$, one value from a range of distances used in the calculations. The partial-annuli in the plot are therefore a sample of much wider annuli.} 
\label{1x3_scheme}
\end{figure}

Before understanding the cause of these variations, it is useful to review some basic geometry used in the far-side imaging measurement schemes.
Helioseismic waves that are used to map the far-side ARs often experience a few surface reflections during their propagation through the Sun. 
An example of the 1$\times$3-lower measurement scheme is shown in Figure~\ref{1x3_scheme}.
Waves starting from the near-side Sun experience a one-skip travel to reach the far-side target, after which the waves continue with a one-skip and a two-skip travel on the far side, and come back to the near side as three-skip waves. 
The oscillatory signals observed in the near-side one-skip and three-skip partial-annuli are used to measure travel-time shifts corresponding to that far-side target. 
In this measurement scheme, the waves travel four skips and touch the surface five times, and each touch of the surface has a possibility of encountering an active region. 
Based on this knowledge of far-side images' measurement processes, we speculate that there may be four possible causes that lead to the measured long-term \dtm-variation.

\begin{enumerate}[i]
\item Far-side ARs manifest themselves as travel-time deficits on the far-side maps. 
With the waxing and waning of the number of ARs on the far side, one would naturally expect that the far-side \dtm\ varies with the solar cycle.

\item Far-side maps are calculated using the near-side Doppler velocities, and the Doppler velocities observed inside the near-side ARs already carry travel-time deficits with them. 
It is thus likely that the travel-time deficits measured in the far-side images come from the near-side oscillatory signals in the near-side ARs, whose travel-time deficits are well anti-correlated with the near-side magnetic activity.

\item As pointed out above, although all of the different measurement schemes focus on the same far-side target, the waves actually get reflected a few times on different surface locations before and after getting reflected at the far-side target (see Figure~\ref{1x3_scheme}a). 
These surface-reflection locations, distributed non-uniformly around the globe, have a probability of residing inside magnetic regions, and travel-time deficits during such surface reflections can be carried along the entire journey of the waves. 
The more magnetic regions, the more likely that surface reflections occur inside a magnetic region, and more travel-time deficits may be measured in \dtm.

\item As interpreted by \cite{gon09}, the \dtm-variation is due to the cyclic variation of the Sun's seismic radius $\mathrm{R_\odot}$, which is also in phase with the Sun's magnetic activity. 

\end{enumerate}

Not all these possible causes are completely independent from each other; for instance, causes 1 and 2 are certainly parts of cause 3. 
But what we want to find out here is the dominant factor that determines the \dtm-variation trend. 
If cause 1 dominates, one would expect that the \dtm\ shows different trends in the northern and southern hemispheres, similar to what we see in the magnetic-field trends; but Figure~\ref{NS_hemi} shows that this is not the case. 
If cause 2 dominates, one would expect that the \dtm\ with a different number of wave skips show similar amount of variations because essentially similar near-side oscillatory signals are used in all of these calculations; however, Figures~\ref{time_central}a and \ref{time_limb}a show that this is unlikely to be the case. 
However, one also needs to be more cautious here, because the waves with different numbers of skips are actually waves of different oscillatory modes $(n, \ell)$, corresponding to waves penetrating to different depths of the Sun. 
This difference likely leads to small discrepancies seen in \dtmn\ in Figures~\ref{time_central}b and \ref{time_limb}b rather than the larger differences seen in \dtm\ in Figures~\ref{time_central}a and \ref{time_limb}a.
Next, we focus on examining which of causes 3 and 4 plays a dominant role in causing the long-term \dtm-variation. 

\subsection{Effect of Surface Magnetic Field or Change of $\mathrm{R_\odot}$?}
\label{sec32}

To assess how the magnetic field, distributed around the global Sun, impacts the \dtm-measurements, as listed as cause 3 above, we need to assess the probability of acoustic waves, which carry far-side information back to the near side, encountering a magnetic region on the solar surface during their journey.
We first assess for each measurement scheme the probability density of a waves' surface reflections as a function of latitude. 
This probability density is also a function of longitude; however, because in this study we only present the 120-day running-average results, the longitude dependency can be omitted. 

Figure~\ref{1x3_scheme}a shows one of the 14 measurement schemes that are used to measure the travel-time shift corresponding to one far-side target, and Figure~\ref{1x3_scheme}b shows one sample of a series of skip distances that are used in this 1$\times$3-lower measurement scheme. 
As one can see, the reflection points of the waves that are used to measure one far-side location are distributed over most of the Sun's latitudes. 
In fact, for one measurement scheme and for one far-side target point, there are tens of thousands of surface reflection locations associated with it even in our data with a low spatial-resolution of $0.8\degr$ pixel$^{-1}$, and the probability density as a function of latitude $\lambda$ [\pdf] is a measure of the distribution of these surface-reflection locations. 
These surface-reflection locations can be found through spherical geometry calculations, and the total number of reflections is then counted for each $5\degr$-wide latitudinal bands from $-90\degr$ to $+90\degr$.
The same computation is repeated for each far-side target for this measurement scheme, and then the same process is repeated for all the 14 measurement schemes.
The \pdf\ can then be found for each latitudinal band by normalizing relative to the total number of reflections across all latitudes.

\begin{figure}[!t]
\centerline{\includegraphics[width=0.60\textwidth]{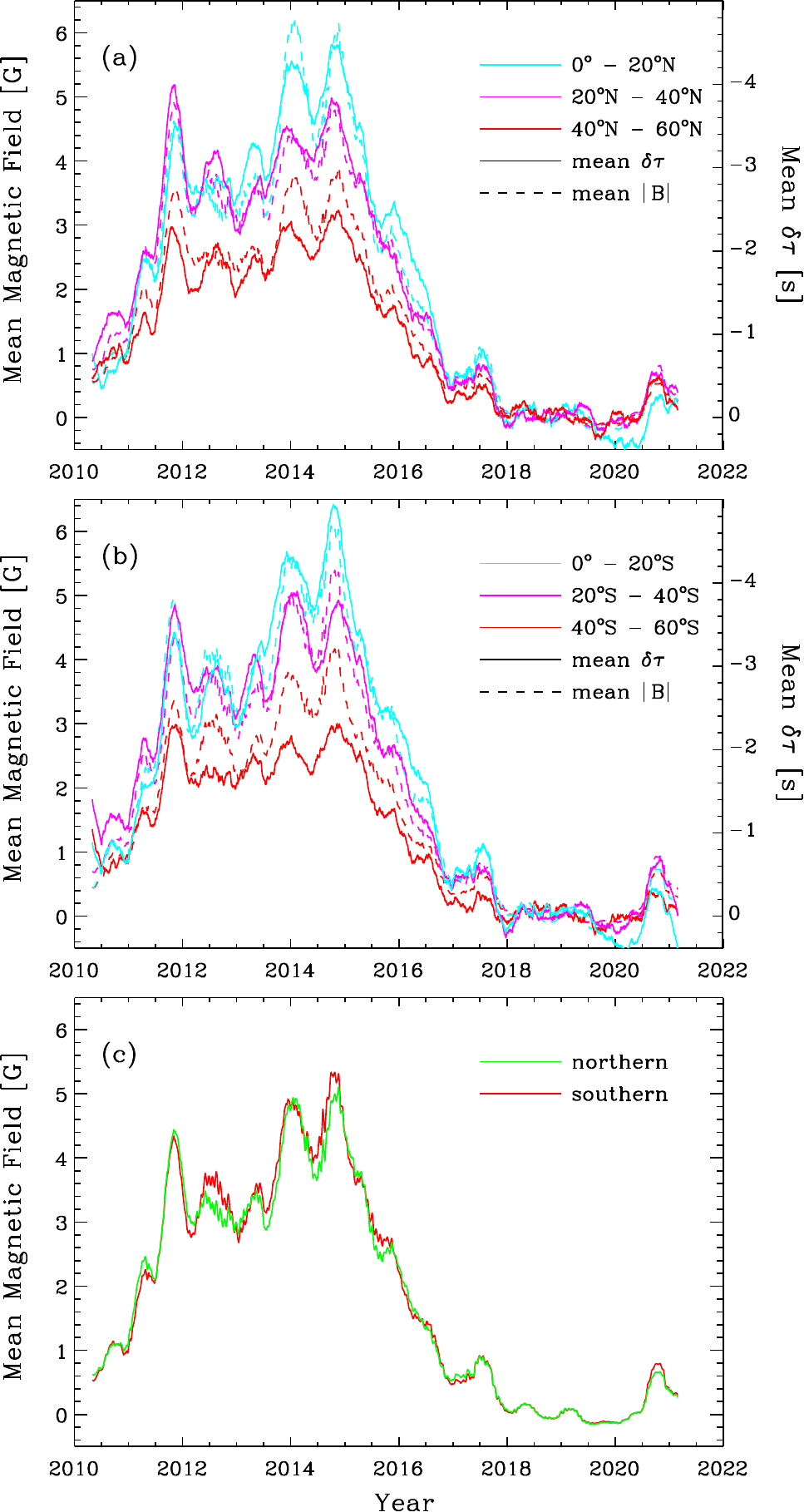} }
\caption{Temporal variations of far-side \dtm\ ({\it solid lines}), plotted decreasing upward, and \bave\ ({\it dashed lines}) for selected latitudinal bands in both northern {\bf (a)} and southern {\bf (b)} hemispheres. 
{\bf (c)} Comparison of the \bave\ integrated in the entire northern and southern hemispheres separately.}
\label{meanT_estM}
\end{figure}

Likewise, for each day, we average the absolute values of the full-disk magnetic field into $5\degr$-wide latitudinal bands, which is $|B(\lambda,t)|$, where $t$ is time (or day in this case). 
We then calculate an averaged magnetic-field strength at all of the waves' reflection locations, i.e. modulated by \pdf\ following $$\langle |B(\lambda, t)| \rangle = \mathrm{pdf}(\lambda) |B(\lambda,t)|.$$
\bave\ measures the averaged magnetic-field strength that the waves encounter on the surface along their entire journey from the near side to the far side and then back to the near side.
These \bave\ are then averaged again into 20$\degr$-wide latitudinal bands for each day to compare against the \dtm-measurements in Figure~\ref{lat_var}, and Panels a and b of Figure~\ref{meanT_estM} show the comparison after a 120-day running average.
For the latitudinal bands of $0\degr - 20\degr$ and $20\degr - 40\degr$ in both hemispheres, the trend and values of \dtm\ and \bave\ match pretty well, implying there is a single scaling factor between these two quantities. 
However, the \dtm\ in the $40\degr - 60\degr$ band are notably smaller than \bave\ with that same scaling factor, and this will be further discussed in Section~\ref{sec33}. 
Figure~\ref{meanT_estM}c shows that the \bave, integrated separately in both the northern and southern hemispheres, are very similar to each other, just as one would expect if comparing with Figure~\ref{NS_hemi}a.

We thus believe that the long-term variations of \dtm\ are caused collectively by the magnetic field distributed around the Sun, whose interaction with the traveling acoustic waves at their surface reflection locations around the Sun leaves a travel-time deficit on the far-side acoustic maps.
We tend to believe that the long-term \dtm\ variation is not caused by the seismic radius $\mathrm{R_\odot}$ variation, because that can hardly explain the latitudinal differences measured in \dtm. 
However, we also want to stress that this does not necessarily indicate that the $\mathrm{R_\odot}$ does not vary with the solar cycle. 
Our analysis only demonstrates that the magnetic field around the Sun, through interacting with the the waves' surface reflections, is collectively the dominant factor that causes the long-term variation measured in \dtm, and what factor the $\mathrm{R_\odot}$ variation plays in the \dtm\ is not studied here, but it is believed not to play a substantial role.

Meanwhile, we also recognize that the long-term variation of the \dtm\ is caused by magnetic field, but is mostly materialized through the Wilson depression associated with magnetic field \citep{lin10} along with the interior thermal changes in and around magnetic regions. 
Averaging local Wilson depressions across all longitudes in indivdual latitudinal bands can result in the latitude-dependent reductions of the solar radius, and this reduction can also in the same sense explain the analysis results shown in Figure~\ref{meanT_estM}.
However, this radius reduction is mostly a local and near-surface phenomenon, not a global-scale isotropic seismic-radius change in the relatively deeper interior. 

\subsection{A Few Discrepancies}
\label{sec33} 

So far, there are a few discrepancies in the \dtm-measurements and in our explanation of measurements that are not negligible. 
These include: i) Comparing Figures~\ref{time_central}b and \ref{time_limb}b, one can find that the \dtmn\ have clear differences in the near-center and near-limb measurements, around $-0.7$\,seconds and $-0.9$\,seconds, respectively, in the solar activity maximum years.
ii) As shown in Figure~\ref{meanT_estM}, the measured \dtm\ in the latitudinal bands of $40\degr - 60\degr$ are substantially smaller than \bave\ after scaling.
iii) In either Figure~\ref{time_central}b or \ref{time_limb}b, \dtmn\ is different for different wave-skip numbers despite measuring the same area of the far side.

The first discrepancy, the \dtmn\ measured in the near-center area differs from that measured in the near-limb area by nearly 0.2\,seconds, is likely due to the helioseismic center-to-limb effect \citep{zha12, che18}. 
The measurements for the far-side central area often uses more observations from the near-side central area than the near-side limb area, while the measurements for the far-side limb area often uses more observations from the near-side limb area than from the near-side central area. 
The helioseismic center-to-limb effect thus leaks into the far-side measurements, causing the \dtm\ measured in the far-side limb area to be relatively smaller.

The second discrepancy, that the \dtm\ and \bave\ do not match well in the high-latitude bands, may be due to the following reasons: 
First, similar to the first discrepancy, the helioseismic center-to-limb effect causes the far-side \dtm\ measurements in higher latitude (limb area) to be substantially smaller than in low-latitude areas (central area).
Second, the magnetic field's effect on the travel-time shift is understandably not linear, and weaker magnetic field in the high-latitude areas may have an even weaker effect on travel-time shifts, resulting in smaller \dtm\ in high latitudes.


The third discrepancy, the \dtmn\ measured from a different number of wave skips, may reflect the real solar-interior variations. 
Because these waves are of different oscillatory modes probing to different depths of the solar interior, as pointed out in Section~\ref{sec31}, the small differences in \dtmn, an order of 0.1\,seconds, probably imply the interior physical changes associated with the solar cycle. 
It is possible that the waves of slightly different modes may have slightly different travel-time shifts when reflected from the same Wilson depression; or it is possible that the solar-interior structure changes more near the surface than in the deeper region so that waves with more skips, i.e. higher-$\ell$ modes, experience slightly longer travel-time shifts. 
These different scenarios are all worth further study.

\section{Conclusion}
\label{sec4}

The time--distance far-side images from 11 years of SDO/HMI observations exhibit a systematic variation in the disk-averaged travel-time shifts, which exhibit a very good anti-correlation with the near-side magnetic-flux density that varies with the solar cycle. Our multiple multi-skip measurement schemes that are used to image the far side of the Sun allow us to analyze the single-skip travel-time deficits and the travel-time deficits measured in different latitudinal bands. Our analysis demonstrates that the long-term variation in the far-side travel-time shifts are primarily caused, collectively, by the surface reflections, which unavoidably interact with the magnetic field that is distributed non-uniformly around the whole Sun, rather than caused by the seismic-radius variation. This analysis does not exclude the possibility that the seismic radius may vary with the solar cycle, but the variation of the seismic radius is not the dominant cause of the long-term variation of the measured far-side travel-time shifts.

\begin{acks}
We thank an anonymous referee for his/her valuable comments and sugguestions that help improve the quality of this article.
\end{acks}

\begin{fundinginformation}
This work was partly funded by the NOAA grant NA18NWS4680082.
\end{fundinginformation}

\begin{dataavailability}
The far-side images used in this analysis are made by the SDO/HMI's time--distance far-side imaging pipeline, and can be downloaded from the webpage: \url{http://jsoc.stanford.edu/data/timed/}. 
The near-side magnetograms are from the SDO/HMI observations and can be downloaded from the webpage: \url{http://hmi.stanford.edu/magnetic/}.
\end{dataavailability}

\begin{conflict}
The authors declare that they have no conflicts of interest.
\end{conflict}

\bibliographystyle{spr-mp-sola}
\bibliography{ms_bib}

\end{article}
\end{document}